\documentclass[aps,nofootinbib,notitlepage,superscriptaddress,twocolumn,10pt,prd]{revtex4-1}

\usepackage{graphicx}
\usepackage{amsmath,amssymb}
\usepackage{hyperref}
\usepackage{braket}
\usepackage{subfigure}
\usepackage{float}
\usepackage[dvipsnames]{xcolor}
\usepackage{soul}
\usepackage{rotating}
\usepackage{multirow}
\usepackage{mathtools}

\usepackage{makecell}

\usepackage[margin=0.75in]{geometry}

\hypersetup{
	colorlinks=true,
	linkcolor=red,
	citecolor=blue,
}

\usepackage{xcolor}

\newcommand{\lens}{l}
\newcommand{\source}{s}

\usepackage{natbib}

\begin{document}
	
\title{A model independent comparison of supernova and strong lensing cosmography: implications for the Hubble constant tension} 

\author{Shivam Pandey}
\affiliation{Center for Particle Cosmology, Department of Physics and Astronomy, University of Pennsylvania, Philadelphia, PA 19104, USA}
\author{Marco Raveri}
\affiliation{Center for Particle Cosmology, Department of Physics and Astronomy, University of Pennsylvania, Philadelphia, PA 19104, USA}
\author{Bhuvnesh Jain}
\affiliation{Center for Particle Cosmology, Department of Physics and Astronomy, University of Pennsylvania, Philadelphia, PA 19104, USA}

\begin{abstract}	
We use supernovae measurements, calibrated by the local determination of the Hubble constant $H_0$ by SH0ES, to interpolate the distance-redshift relation using Gaussian process regression. We then predict, independent of the cosmological model, the distances that are measured with strong lensing time delays.
We find excellent agreement between these predictions and the  measurements. The agreement holds when we consider only the redshift dependence of the distance-redshift relation, independent of the value of $H_0$. Our results disfavor the possibility that lens mass modeling contributes a 10\% bias or uncertainty in the strong lensing analysis, as suggested recently in the literature. 
In general our analysis strengthens the case that residual systematic errors in both measurements are below the level of the current discrepancy with the CMB determination of $H_0$, and supports the possibility of new physical phenomena on cosmological scales. 
With additional data our methodology can provide more stringent tests of unaccounted for systematics in the  determinations of the distance-redshift relation in the late universe.  
\end{abstract}
	
\maketitle

Direct measurements of the Hubble constant, $H_0$, and its inference from the Cosmic Microwave Background (CMB) are showing a discrepancy of increasing statistical significance within the $\Lambda$CDM cosmological model (see~\cite{Verde:2019ivm} for a recent review).

Measurements of $H_0$ based on supernovae, calibrated with cepheid variables~\cite{Riess:2019cxk}, and the results of CMB observations from the {\it Planck}  satellite~\cite{Aghanim:2018eyx}  differ at a confidence level equivalent to $4.4$ standard deviations, ruling out the possibility that this discrepancy is due to a statistical fluctuation.

The challenging aspect of this tension is to understand if it is related to residual unaccounted systematic effects in the data or to a breakdown of the standard cosmological model that could result in the discovery of new physical phenomena on cosmological scales.

In this context, distance measurements from strong lensing time delays, as reported recently in~\cite{Wong:2019kwg}, are crucial as they do not share any source of systematic uncertainty with the other two probes.
These use the variability of quasars lensed into multiple images by elliptical galaxies to estimate the time delay along the different lines of sight. By fitting for a model for the lens mass distribution, they constrain a combination of lens and source distances as detailed below. 
Sources and lenses are separated by cosmological distances so that the measured combination depends on both the amplitude and shape of the distance-redshift relation.

In this letter we carry out a consistency test of the distance-redshift relation measured by the supernovae in the Pantheon compilation (SN)~\cite{Scolnic:2017caz}  and the available time delay and lens distance measurements from H0LiCOW~\cite{Suyu:2016qxx,Wong:2019kwg}. Our tests span  all redshifts that are probed and is independent of the underlying cosmological model. We carry out tests with the SN distances calibrated with local measurements of the Hubble constant by SH0ES~\cite{Riess:2019cxk} as well as tests that are free of the $H_0$ calibration. 
We leverage  the large number of measured SN distances  to reliably interpolate between them, using Gaussian Process (GP) regression, and then predict the time delay and lens distances that should be observed. 
Since the two data sets that we consider do not share the same sources of systematic uncertainties this provides an end to end test that is sensitive to residual problems with either dataset. In addition to testing for a bias in the distance-redshift relation, we also test for under-estimated uncertainties by comparing the scatter in the predicted and measured distances. 

Our tests compare SN and strong lensing measurements and are thus not sensitive to the possibility that both experiments have residual systematic bias affecting both of them in the same way.
Although we cannot exclude this possibility, systematic biases that preserve the agreement, across redshift, of both probes seem unlikely. 
We also test the consistency of both, SN and H0LiCOW with Baryon Acoustic Oscillation (BAO) measurements in a model independent way.

These tests are complementary to other consistency tests between SN and H0LiCOW that have recently been discussed in literature~\cite{Taubenberger:2019qna,Arendse:2019hev, Liao:2019qoc} and to  studies probing the sources of systematic effects in the data sets considered here~\cite{Rigault:2014kaa, Rigault:2018ffm, Jones:2018vbn,Schneider:2013wga,Xu:2015dra, Unruh:2016adf,Sonnenfeld:2017dca,Kochanek:2019ruu, Camarena:2019moy}. 

\begin{figure*}[tbp]
\centering
\includegraphics[width=\textwidth]{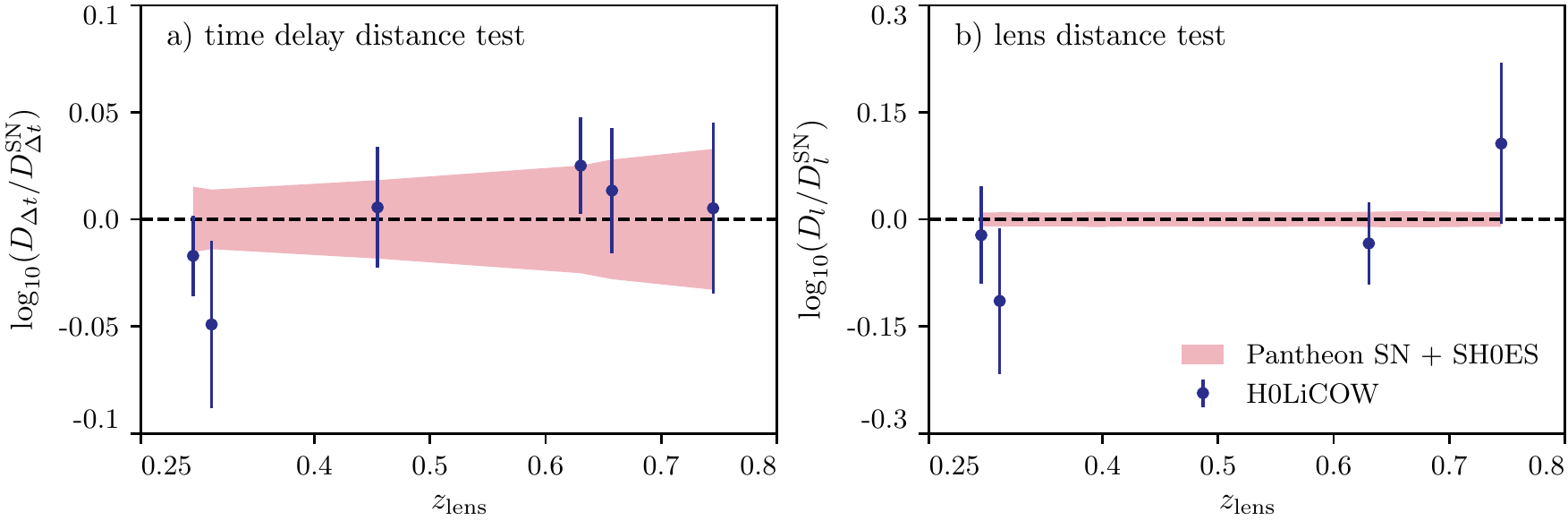}
\caption{ \label{Fig:DistanceComparison}
Comparison of the H0LiCOW measured time delay distance $D_{\Delta t}$ (a) and lens distance $D_\lens$ (b) with the calibrated Pantheon SN predicted values $D_{\Delta t}^{\rm SN}$ and $D_\lens^{\rm SN}$. The points with error bars show the strong lensing estimates while shaded regions indicate the uncertainty in the SN predictions. 
The time delay distances in Panel (a) agree at the $85.7\%$ confidence level.
The lens distances in Panel (b) agree at the $63.6\%$ confidence level. 
An important implication of this test is that the scatter in the H0LiCOW  distances is consistent with the reported errors in the SN and time delay distances.
}
\end{figure*} 
\section*{Data and methodology}
We start by considering the six strong lensing time delay measurements, of which five were analyzed blindly, from the H0LiCOW collaboration in~\cite{Wong:2019kwg,Suyu:2009by,Suyu:2013kha,Wong:2016dpo,Birrer:2018vtm,Rusu:2019xrq,Chen:2019ejq}. The difference of excess time delays between two lensed images (and angular positions $\mathbf{\theta}_i$ and $\mathbf{\theta}_j$) of a source (at angular position $\mathbf{\beta}$) is given by:
\begin{align} \label{Eq:TimeDelay}
\Delta t_{ij} = \frac{D_{\Delta t}}{c} \Bigg[ \frac{(\mathbf{\theta}_i - \mathbf{\beta})^2}{2} - \psi (\mathbf{\theta}_i) - \frac{(\mathbf{\theta}_j - \mathbf{\beta})^2}{2} + \psi (\mathbf{\theta}_j) \Bigg] \,,
\end{align}
where, $\psi(\mathbf{\theta})$ is the lens potential. By measuring time delays and modelling the gravitational potential of the source and the lens one can infer the time delay distance:
\begin{align} \label{Eq:TimeDelayDistance}
D_{\Delta t} \equiv (1+z_\lens)\frac{D_\lens D_\source}{D_{\lens\source}} \,,
\end{align}
where, hereafter, subscripts $\lens$ and $\source$ indicates quantities referring to the lens and source respectively and $D$ is the angular diameter distance of different objects.
As we can see, the measured time delay distance is sensitive to the expansion history of the universe through its dependence on the angular diameter distance at two different redshifts.
For the measured systems the spread in redshift is large, with the lens redshift ranging in $z\in [0.3,0.7]$ and the source redshifts ranging in $z\in [0.6,1.7]$ thus making the time delay distance measurements sensitive to the shape of the distance-redshift relation.

In addition to the time delay distance measurements, lens kinematic data can be used to estimate the angular diameter distance of the lens, $D_\lens$, by comparing the dynamical mass with the lensing mass (which depends on distances) ~\cite{Paraficz:2009xj,Chen:2019ejq, Jee:2014uxa, Birrer:2015fsm, Jee:2019hah}.

The measurements of $D_{\Delta t}$ and $D_\lens$ are in general correlated since they use different aspects of the same data \cite{Birrer:2015fsm, Birrer:2018vtm}.
Since the H0LiCOW collaboration has not yet publicly released the full posterior of the two distance measurements for all but one of its observations, we separately consider the marginalized constraints on $D_{\Delta t}$ and $D_\lens$. 

We observe that, if we consider logarithmic distances, $\log_{10}D_{\Delta t}$ and $\log_{10}D_\lens$, then the posterior distribution of the H0LiCOW measurements becomes practically indistinguishable from Gaussian.
The reason why logarithmic distances are likely to be Gaussian distributed, or very close to that, is that they are computed as the difference of well measured quantities rather than their ratios. 
We discuss further details on the treatment of the strong lensing measurements in Appendix~\ref{App:SLGaussianTest}.

This allows us to convert the constraints on $D_{\Delta t}$ and $D_\lens$ from~\cite{Wong:2019kwg} into constraints on $\log_{10}D_{\Delta t}$ and $\log_{10}D_\lens$, properly accounting for the Jacobian of the transformation, and to consider the latter to be Gaussian distributed.
We  check that cosmological results obtained by fitting the logarithmic measurements reproduce the ones reported in~\cite{Wong:2019kwg}, as detailed in Appendix~\ref{App:SLGaussianTest}.

We then consider the Pantheon SN compilation~\cite{Scolnic:2017caz} that provides accurate measurements of relative distances across the redshift range $z\in [0.01,2.26]$ with $1048$ SN measurements.
We use the measurement of the Hubble constant of $H_0=74.03\pm 1.42$ from the  SH0ES project~\cite{Riess:2019cxk}. Cepheid variables are used to calibrate the absolute magnitude of the SN so that the measured distance modulus can be used to directly estimate luminosity distances.
Further details on the calibration of the SN distance modulus can be found in Appendix~\ref{App:SN_cal}. While the SH0ES analysis is the most mature and precise  of the local measurements of $H_0$,  analyses with alternative approaches are underway.  A recent analysis using the Tip of the Red Giant Branch method by the Carnegie-Chicago Hubble Program~\citep{Freedman:2019jwv} yields a lower value of $H_0$ with somewhat larger uncertainty than SH0ES (but there is some debate about their analysis, see~\cite{Yuan:2019npk}).
We test  this alternative result also in Appendix~\ref{App:SL_Bias}.
Note that the SN sample cannot be calibrated using the $H_0$ determination from CMB measurements in a model independent way. 
Not surprisingly, when using the standard cosmological model determination of the sound horizon scale, ~\cite{Macaulay:2018fxi,Alam:2016hwk,Aubourg:2014yra} find $H_0$ consistent with the CMB value. 
Finally we note that bias corrections of SN luminosities assume the $\Lambda$CDM model and in principle this breaks model independence, but as discussed in~\cite{Kessler:2016uwi} it is a very small effect.

Given that the redshift range spanned by SN observations is populated by a large number of measurements we can interpolate between them to obtain an estimate of distances as a function of redshift.
This is achieved by Gaussian process regression of the measured distance modulus.
The Gaussian process kernel and kernel parameters are chosen so that the Gaussian process inference is as close as possible to the binned Pantheon SN sample.
We find that this procedure is flexible enough to fully capture all the features that are present in the binned SN sample, starting from the full sample, thus effectively obtaining a version of the binned SN sample that is continuous in redshift. We also test the GP covariance matrix, and compare it to the result from polynomial interpolation which tends to significantly  underestimate the error bars at intermediate redshifts. 
Further details on the implementation of the Gaussian process regression are discussed in Appendix~\ref{App:SN_GP}. 

\begin{figure*}[tbp]
\centering
\includegraphics[width=\textwidth]{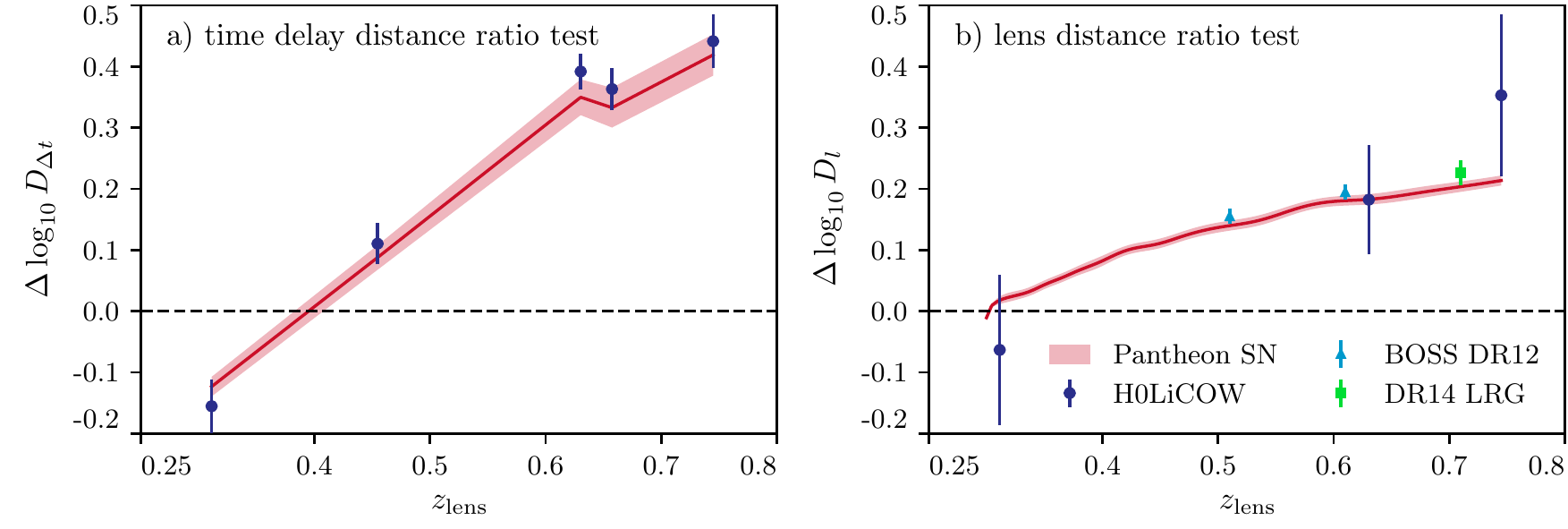}
\caption{\label{Fig:DistanceRatioComparison}
Comparison of the H0LiCOW measured time delay distance ratios (a) and lens distance ratios (b) with the Pantheon SN predicted values (shaded bands). 
 The ratio of angular diameter distances predicted from BAO are also shown in Panel b).
The time delay distances in Panel a) agree at the $78.2\%$ confidence level
while the lens distances in Panel b) agree at the $54.5\%$ confidence level.
The three BAO points in Panel b) and predictions from Pantheon SN agree at $95.4\%$ confidence level.
Note that since only distance ratios are considered in this test, the results are independent of the value of $H_0$. 
}
\end{figure*} 

With the Gaussian process for the SN we can now predict, with Eq.~\eqref{Eq:TimeDelayDistance}, the time delay and lens distances that H0LiCOW should have observed, independently of the cosmological model. We note that the predicuted distances from SN are correlated; we take this into account when computing the statistical significance of the reported results.

\section*{Results}
In Fig.~\ref{Fig:DistanceComparison} we show the results of the SN prediction for the time delay and lens distance measurements.
In Panel a) we can see that the time delay distance measurement predicted by SN agrees with the direct measurements from H0LiCOW.
The uncertainties in the two methods are comparable.
In Panel b)  the lens distance as directly measured by H0LiCOW and predicted by SN are compared.
As in the previous case these are largely in agreement.
The error bars of the SN predicted distances are significantly smaller than the H0LiCOW ones because this quantity is directly measured by a large number of SN at intermediate redshifts.

To precisely quantify the agreement between the H0LiCOW measurements and the SN ones, we exploit the fact that both logarithmic distances are close to Gaussian distributed.
The difference between the two, weighted by the inverse sum of the two covariances is then chi squared distributed with number of degrees of freedom equal to the number of data points.
This results in a probability of agreement of $85.7\%$ for the time delay distances and $63.6\%$ for the lens distance.
Both results indicate very good agreement.

Further, in Fig.~\ref{Fig:DistanceComparison}, there are no outlier measurements.
Since the two distributions are very close, possible non-Gaussianities in the tails of the distributions are not expected to change the results significantly. This also rules out the possibility of a systematic uncertainty in the strong lensing sample at a level larger than the reported error bars.

Beyond the above tests, it is crucial to test whether the SN and strong lensing measurements agree on the amplitude and ``shape'' (the redshift dependence) of the distance-redshift relation. 
The amplitude test relates to possible discrepancies in the determination of the Hubble constant, while being fully independent of the expansion history.
The shape test would tell us if there is agreement between the two measurements regardless of the overall calibration that measures the Hubble constant. 

The amplitude of the distance-redshift relation can be tested by looking at the average residual logarithmic distance.
The H0LiCOW data measure the average amplitude to be $\langle \log_{10}D_{\Delta t} \rangle = 3.514 \pm 0.012$ while the SN predict the average amplitude to be $\langle \log_{10}D_{\Delta t}^{\rm SN} \rangle = 3.517 \pm 0.017$. These two determinations are in full agreement at $89\%$ confidence level and provide a test at about $5\%$ precision. 
For lens distances we similarly have $\langle \log_{10}D_{\Delta t} \rangle = 3.026 \pm 0.044$ and $\langle \log_{10}D_{\Delta t}^{\rm SN} \rangle = 3.0416 \pm 0.0088$ which are again in full agreement at about $10\%$ precision.

To test the redshift dependence of the measurements we can consider distance ratios, or differences in logarithmic distances.
Since the Hubble constant enters as an overall distance multiplier, its value  cancels in the ratio.
We then take both data sets and consider the ratio with respect to the system with the lowest lens redshift.
Although arbitrary, this choice does not influence the outcome of the test. All possible choices would just be different linear combinations of the same data thus leaving the statistical significance unchanged.

We show in Fig.~\ref{Fig:DistanceRatioComparison} the comparison of distance ratios, as predicted by SN measurements and as estimated by H0LiCOW.
As in Fig.~\ref{Fig:DistanceComparison}, the ratios also show agreement between the two measurements.
And since logarithmic distance differences are also close to Gaussian distributed, we can easily compute the statistical significance of their agreement. 
The time delay distance ratios are in agreement at the $78.2\%$ level and the lens distance ratios are in agreement at the $54.5\%$ confidence level.
We note that the H0LiCOW measurements are incompatible with no redshift dependence and hence detect the shape of the distance-redshift relation at $34\sigma$ in Panel a) and $2.6\sigma$ in Panel b).

We can further compare these distance ratios with predictions from the BAO measurements. The BAO measurements are sensitive to the ratio of angular diameter distance of the galaxies and the photo-baryon sound horizon scale at the epoch of decoupling. 
Therefore, we can take the ratio of BAO measurements at two different redshifts and remove any sensitivity to the overall calibration with the sound horizon.
In this analysis we consider three BAO measurements from BOSS DR12 data~\cite{Alam:2016hwk} and one measurement from the eBOSS DR14 LRG data~\cite{Bautista:2017wwp}. We use the lowest redshift DR12 measurement as our reference BAO measurement in the distance ratio test. 
The comparison between the BAO and SN predicted distance ratios can be performed as before and results are in agreement at the 95.4\% confidence level. To qualitatively compare the results to the H0LiCOW measurements we shift the overall amplitude to match the H0LiCOW reference redshift. We compare the distance ratios of all three probes in Fig.~\ref{Fig:DistanceRatioComparison}. 
All measurements show remarkable consistency in this test, which is  independent of $H_0$ and of the cosmological model.

Moreover, if we use the SN Gaussian process to calibrate the BAO angular diameter distances, as discussed in~\cite{Aylor:2018drw}, we can directly measure the sound horizon scale independently of the cosmological model.
Individually the four calibrated BAO measurements that we consider are in good agreement on the sound horizon determination, giving: $137.39 \pm 3.91$ Mpc, $136.05 \pm 3.83$ Mpc and $137.48 \pm 3.92$ Mpc for the three SDSS DR12 observations in increasing order of their effective redshift; $133.77 \pm 6.24$ Mpc for the SDSS DR14 LRG observation.
Jointly the four measurements result in a sound horizon measurement of $135.92 \pm 3.26$ Mpc, which is in $3.3 \sigma$ tension with the {\it Planck} results of $147.09 \pm 0.26 $ Mpc~\cite{Aghanim:2018eyx}.
This effectively accounts for a large portion of the $4.4 \sigma$ Hubble constant tension. 
Since the sound horizon is constant after recombination, this part of the Hubble constant tension cannot be resolved by introducing new physics between the redshift of the BAO measurements and recombination.

\section*{Summary and discussion}
We have presented a new way of testing the consistency of distance measurements from supernovae and strong lensing time delays that is independent of the cosmological model.
Our method exploits the power of SN observations across a large range of redshifts to directly predict the outcome of strong lensing measurements. We use Gaussian process regression to interpolate across redshift and show that these provide robust results. Tests of the covariance matrix show that it is more reliable than polynomial interpolation which tends to underestimate the error bars at intermediate redshifts. 

We devise three types of tests sensitive to the distance-redshift relation: one directly tests distances (and therefore biases in $H_0$), one tests  their calibration and another tests distance ratios. 
While the first test is sensitive to both calibration and redshift dependent systematic effects, the other two single out and test these two aspects independently.

We find that all tests report excellent agreement between the Pantheon supernovae, calibrated with the SH0ES distance ladder, and H0LiCOW time delay measurements. 
Given the model independence of our tests we conclude that, at present sensitivity, 
there is no indication for the presence of unaccounted systematic effects in either data set. In particular, if the distance-redshift relation inferred from SN is correct, there is no evidence of a residual bias due to mass modeling uncertainties in the strong lensing data. It is possible that both measurements have a bias of the same sign, magnitude and redshift dependence -- this unlikely scenario would evade our tests. 

The possibility that uncertainties in mass modeling, in particular the mass sheet degeneracy, have biased the strong lensing determination of $H_0$ has been discussed in the literature \citep{Schneider:2013wga,Xu:2015dra, Unruh:2016adf,Sonnenfeld:2017dca,Kochanek:2019ruu}.
Recently~\cite{Kochanek:2019ruu} has suggested that this leads to at least a 10\%  level of uncertainty on $H_0$ in present and  near future measurements. 
We explicitly check the impact  an unknown residual systematic would have on both $D_{\Delta t}$ and $D_{\lens}$. We create fake strong lensing observations and test for several types of unaccounted for errors: 
\begin{itemize}
    \item Biases in the distance-redshift relation that directly impact  $H_0$ or the amplitude of the distance-redshift relation. An 8\% bias that would fully reconcile the tension with {\it Planck} is disfavored at nearly 3$\sigma$. 
    \item Redshift dependent biases. We test the ratio of distances as well as a bias that scales as $(1+z)$, motivated by uncertainties in lens mass modeling. These are also disfavored at over 2$\sigma$. 
    \item Underestimated errors. We artificially increase the covariance in the lensing measurements to obtain 10\% uncertainty on $H_0$. We then find that the probability of disagreement with SN inferred distances drops to 0.05\%, i.e. over 3$\sigma$ evidence against inflated errors. 
\end{itemize}

Unknown measurement systematic effects and/or incorrect lens modelling are likely to affect different systems differently, so they  would show up as an unexpected redshift dependence of the measurements. On the other hand if they affected all lenses the same way, they would affect the amplitude. 
These problems are not guaranteed to show up as a discrepancy in the determination of some cosmological parameter, so our tests provide an independent check. We discuss the detailed results of these tests in Appendix~\ref{App:SL_Bias}. 

Strong lensing and SN determination of distances almost certainly do not share the same systematic uncertainties.  The consistency we have found between the two sets of measurements implies that the discrepancy with the CMB is more likely due to new physical phenomena, or a potential systematic error in the CMB analysis (which is considered unlikely as it has been tested extensively~\cite{Addison:2015wyg,Aghanim:2016sns,Hou:2017smn,Aylor:2017haa,Aghanim:2019ame}). 
With reduced uncertainties, as expected by increases in the number of SN~\cite{Abbott:2018wog, Scolnic:2019apa, Hlozek:2019vjs,Shajib:2019toy} and new lens systems~\cite{Huber:2019ljb, Yildirim:2019vlv}, we expect our methodology to provide more stringent tests for systematic uncertainties. In the near term joint analysis of the full $D_{\Delta t}$ and $D_{\lens}$ posterior will also strengthen our tests. 
With a factor of two reduction in statistical uncertainties, we could rule out a 5 percent bias at the 3$\sigma$ level, independent of the cosmological model -- this would certainly strengthen the case for new physics. 

We find that measurements of BAO distances are also in agreement with predictions from SN.  Once calibrated with SN, the BAO measurements are in tension with the CMB over the determination of the sound horizon. 
We find that this accounts for a large part of the statistical significance of the Hubble constant tension.
Since the sound horizon is constant after recombination, this tension is largely independent of the expansion history between the redshift of the measured BAOs, of about $z\sim 0.7$, 
and recombination.
The structure of the CMB peaks measures the angular size of the sound horizon very precisely. Thus an explanation of the Hubble constant tension that would significantly change the inference of the CMB parameters must rely on new physical phenomena before recombination (see~\cite{Knox:2019rjx} and references therein).

We implicitly assumed in our analysis that the universe is spatially flat. It is possible to perform the same tests assuming a curved universe. Since this will introduce an extra free parameter our tests would show stronger consistency. 
We defer an analysis of time delay distances and their implications for curvature to a future study (see \cite{Collett:2019hrr, Arendse:2019hev} for related analyses). 

\acknowledgments
We thank 
Eric Baxter, 
Gary Bernstein, 
Simon Birrer,
Dillon Brout, 
Neal Dalal, 
Wayne Hu,
Mike Jarvis and 
Sherry Suyu
for helpful discussions and comments on the paper.
SP is supported in part by the U.S. National Science Foundation award AST-1440226.
MR and BJ are supported in part by NASA ATP Grant No. NNH17ZDA001N, and by funds provided by the Center for Particle Cosmology. 
BJ is supported in part by the US Department of Energy Grant No. DE-SC0007901. 
Computing resources were provided by the University of Chicago Research Computing Center through the Kavli Institute for Cosmological Physics at the University of Chicago. 

\appendix

\section{The Supernovae calibration with the Hubble constant}\label{App:SN_cal}

In this appendix we discuss the details of the SN distance modulus calibration with measurements of the Hubble constant.
In doing so we mostly follow~\cite{MRinprep}.
The only difference with~\cite{MRinprep} is that, in this work, we add directly the calibration to the SN data points and add its contribution to the SN covariance instead of using an additional parameter to describe the calibration.
We have verified explicitly that these two approaches are equivalent.

The Pantheon collaboration provides measurements of
the SN magnitude relative to a fiducial and arbitrary absolute magnitude, $m-M_{\rm fid}$ with its covariance.
The Pantheon likelihood is then usually marginalized over the value of the fiducial magnitude so that the SN catalog is only
using relative SN distances to constrain cosmological parameters.

The SN distance modulus, $m-M$ is related to luminosity distances, $D_L$, by
\begin{align} \label{Eq:DistanceModulus}
m-M = 5 \log_{10} \frac{D_L(z)}{10\,{\rm pc}} \,.
\end{align}
To use this distance modulus to measure distances we then clearly need to fix the difference between the fiducial and absolute SN magnitude.

Calibrating the absolute magnitude of the very low redshift SN
is precisely the way in which the Hubble constant in~\cite{Riess:2019cxk} is measured. We can then use that determination of the Hubble constant to fix the Pantheon absolute magnitude.

The Pantheon fiducial absolute magnitude, $M_{\rm fid}$, is chosen to correspond to $H_0=70$~\cite{Kessler:2009yy}, quoted here and throughout in units of km\,s$^{-1}$\,Mpc$^{-1}$.

The difference between the fiducial and measured absolute magnitude is then given by:
\begin{align} \label{Eq:SNCalibration}
M-M_{\rm fid} = 5 \log_{10}\frac{H_0}{H_0^{\rm fid}} \,.
\end{align}
The covariance of the calibrated distance modulus has to take into account the uncertainty of the determination of the SN absolute magnitude so that:
\begin{align} \label{Eq:SNCalibratedCovariance}
{\rm cov}(m-M)_{ij} = {\rm cov}(m-M_{\rm fid})_{ij} +\left( \frac{5}{\ln 10} \frac{\sigma_{H_0}}{H_0}\right)^2 \,,
\end{align}
where the indexes $i,j$ run over all the SN in the catalog.

The fully calibrated SN compilation is then sensitive to both the amplitude and the shape of the distance-redshift relation.

We check our implementation by fitting the resulting measurements with the flat $\Lambda$CDM model in which the luminosity distance, as a function of redshift, is given by:
\begin{align} \label{Eq:LCDMLuminosityDistance}
D_L(z) = \frac{1+z}{H_0} \int_0^z\frac{dz'}{\sqrt{\Omega_m (1+z')^3 + 1-\Omega_m}} \,,
\end{align}
where $\Omega_m$ denotes the total matter density in units of critical density today.

\begin{figure}[htbp]
\centering
\includegraphics[width=\columnwidth]{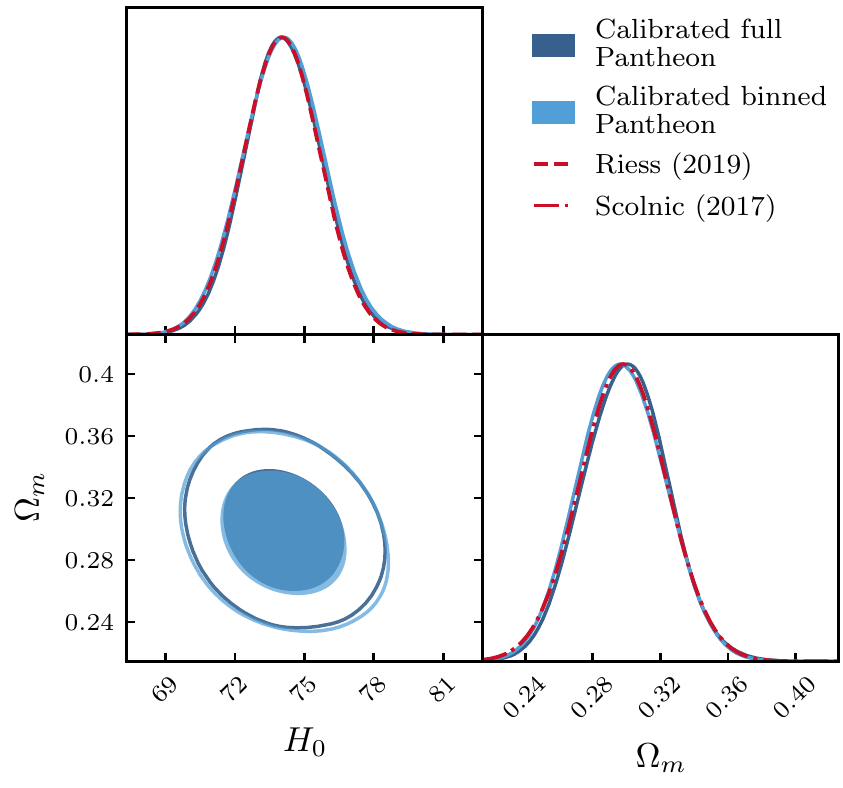}
\caption{ \label{Fig:SNParameterTest}
The marginalized posterior of the $\Lambda$CDM parameters from the calibrated full and binned Pantheon SN sample.
The filled contour corresponds to the 68\% C.L. region while the continuous contour shows the 95\% C.L. region.
The two dashed lines show the constraints on the two cosmological parameters separately, as obtained in~\cite{Riess:2019cxk} for $H_0$ and~\cite{Scolnic:2017caz} for $\Omega_m$ respectively.
We do not show them in the joint contour panel for visual clarity.}
\end{figure}

In Fig.~\ref{Fig:SNParameterTest} we show the joint posterior of $H_0$ and $\Omega_m$ as obtained by fitting both the full and binned calibrated Pantheon SN catalog.
The figure also shows the results reported in~\cite{Riess:2019cxk} and~\cite{Scolnic:2017caz} for the two flat $\Lambda$CDM parameters respectively, showing that our procedure is fully consistent.

We further notice that, in order to directly infer distances from the SN distance modulus, we have to apply a correction that takes into account the difference between redshifts in the heliocentric, $z_{\rm h}$, and CMB, $z_{\rm CMB}$, reference frames.
The luminosity distance at a given redshift in the CMB frame is then given by:
\begin{align}
5\log_{10}\frac{D_L(z_{\rm CMB})}{10\,{\rm pc}} = m-M -5\log_{10}\frac{(1+z_{\rm h})}{(1+z_{\rm CMB})} \,.
\end{align}
We observe that this correction is overall very small.

\section{Distribution of measured time delay distances}\label{App:SLGaussianTest}

In this appendix we discuss the Gaussianity of the measured time delay and lens distance measurements, detailing the analysis of the public posterior distribution of the SDSS 1206+433 system by the H0LiCOW collaboration.
The estimated distances are obtained as ratios of well measured quantities that are  close to Gaussian distributed, but their ratios are not Gaussian distributed.
Logarithmic distances, on the other hand, are sums and differences of well measured quantities and should have a distribution that is closer to Gaussian.
\begin{figure}[htbp]
\centering
\includegraphics[width=\columnwidth]{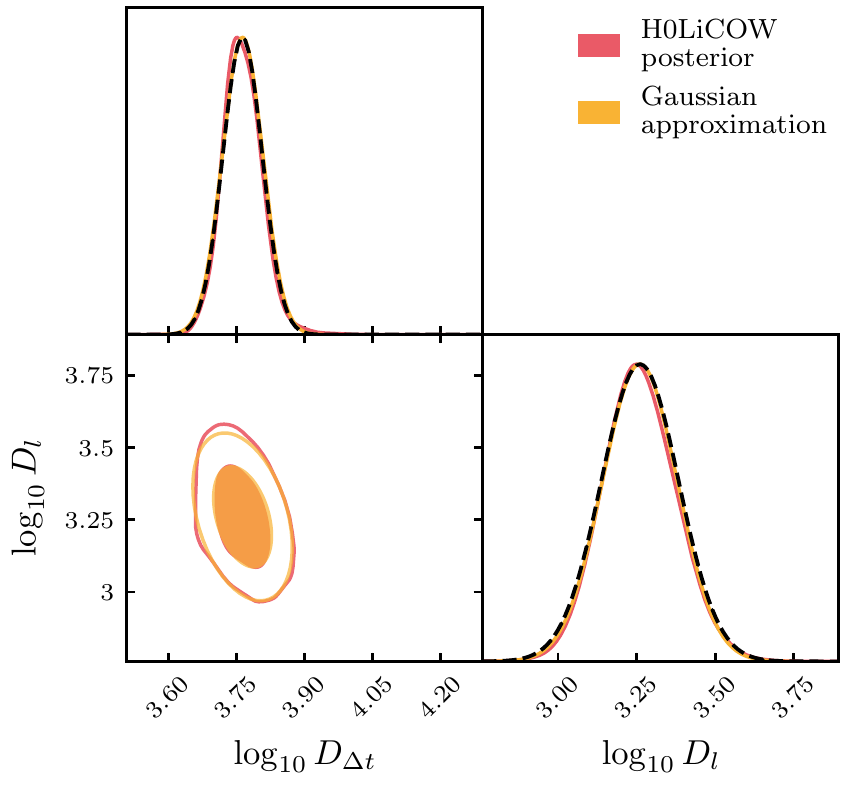}
\caption{ \label{Fig:SLGaussianPosterior}
The joint posterior of logarithmic time delay and lens distances for the SDSS 1206+4332 strong lensing system.
The filled contour corresponds to the 68\% C.L. region while the continuous contour shows the 95\% C.L. region.
The black dashed lines show the posterior of the two separate logarithmic distances as inferred from Eq.~\eqref{Eq:SLLogGaussianization}.
}
\end{figure} 
In Fig.~\ref{Fig:SLGaussianPosterior} we show the posterior of logarithmic time delay and lens distances as obtained form the publicly available posterior samples of the SDSS 1206+433 system, together with its Gaussian approximation.
The full posterior is very close to the Gaussian approximation. 
The 95\% C.L. region shows slight deviations from Gaussianity that are hard to tell apart from sampling noise.

We can then proceed and derive the Gaussian approximation for the two separate distance measurements under the logarithmic distribution assumption.
The H0LiCOW collaboration reports the mean distance together with the $84\%$ and $16\%$ confidence level constraints, in Table 2 of~\cite{Wong:2019kwg}.
Given two constraints, $v_1$ and $v_2$, at two different confidence levels, $p_1$ and $p_2$, for a random variable $X$, assuming that $\log X$ is Gaussian distributed, we can compute the mean, $\langle \log X \rangle$ and variance, ${\rm var}(\log X)$, of $\log X$ as:
\begin{align} \label{Eq:SLLogGaussianization}
\langle \log X \rangle =& \frac{-{\rm Erf}^{-1}(2 p_2-1)|\log v_1 -\log v_2|}{|{\rm Erf}^{-1}(2 p_2-1) -{\rm Erf}^{-1}(2 p_1-1)|} \,, \nonumber \\
{\rm var}(\log X) =& \frac{|\log v_1 -\log v_2|}{\sqrt{2}|{\rm Erf}^{-1}(2 p_1-1) -{\rm Erf}^{-1}(2 p_2-1)|} \,,
\end{align}
where ${\rm Erf}^{-1}$ denotes the inverse error function.
This transformation  takes into account the Jacobian transformation of the two variables.

\begin{table*}[ht!]
\begin{ruledtabular}
\begin{tabular}{ l c c c c c c c }
\textrm{Lens name} & $z_{\lens}$ & $D_{\Delta t}$ (Mpc) & $\log_{10} D_{\Delta t}$ & $\log_{10} D^{\rm SN}_{\Delta t}$ & $D_{\lens}$ (Mpc) &  $\log_{10} D_{\lens}$ & $\log_{10} D^{\rm SN}_{\lens}$ \\
\hline \hline
B1608+656 \cite{Suyu:2009by} & 0.630     & $5156^{+296}_{-236}$ & $3.714\pm 0.022$ & $3.689 \pm 0.025$ & $1228^{+177}_{-151}$ &  $3.090\pm 0.058$ &  $3.123 \pm 0.010$ \\
RXJ1131-1231 \cite{Suyu:2013kha,Chen:2019ejq} & 0.295   & $2096^{+98}_{-83}$   & $3.323\pm 0.019$ & $3.339 \pm 0.015$ & $804^{+141}_{-112}$  &  $2.908\pm 0.068$ &  $2.929 \pm 0.009$ \\
HE 0435-1223 \cite{Wong:2016dpo, Chen:2019ejq} & 0.454   & $2707^{+183}_{-168}$ & $3.433\pm 0.028$ & $3.427 \pm 0.018$ & $-$ & $-$ & $-$ \\
SDSS 1206+4332 \cite{Birrer:2018vtm} & 0.745 & $5769^{+589}_{-471}$ & $3.764\pm 0.040$ & $3.758 \pm 0.032$ & $1805^{+555}_{-398}$ &  $3.261\pm 0.113$ &  $3.154 \pm 0.010$ \\
WFI2033-472 \cite{Rusu:2019xrq} & 0.657   & $4784^{+399}_{-248}$ & $3.686\pm 0.029$ & $3.672 \pm 0.028$ & $-$ & $-$ & $-$ \\
PG 1115+080 \cite{Chen:2019ejq} & 0.311    & $1470^{+137}_{-127}$ & $3.167\pm 0.039$  & $3.216 \pm 0.014$  & $697^{+186}_{-144}$  &  $2.844\pm 0.102$ &  $2.958 \pm 0.010$ \\
\end{tabular}
\end{ruledtabular}
\caption{\label{Tab:SLGaussianResults}
The measured time delay and lens distances and  uncertainties,  along with their predicted values from SN data  (indicated by  the superscript).
The third and sixth columns are from~\cite{Wong:2019kwg}, the fourth and seventh are computed with Eq.~\eqref{Eq:SLLogGaussianization}, while the fifth and eighth are computed with the SN Gaussian process described in Appendix~\ref{App:SN_GP}.
}
\end{table*}
We report in Table~\ref{Tab:SLGaussianResults} the results of computing the logarithmic time delay distances from the measurements in~\cite{Wong:2019kwg}.
We observe that these results are in good agreement with the log-normal likelihoods reported for three of the lenses in~\cite{Suyu:2009by,Suyu:2013kha,Wong:2016dpo}.
Moreover, we show the logarithmic distance posterior derived from the results in Table~\ref{Tab:SLGaussianResults} for the SDSS 1206+4332 strong lensing system. This is indistinguishable, for all practical purposes, from the Gaussian approximation and the full posterior.
\begin{figure}[htbp]
\centering
\includegraphics[width=\columnwidth]{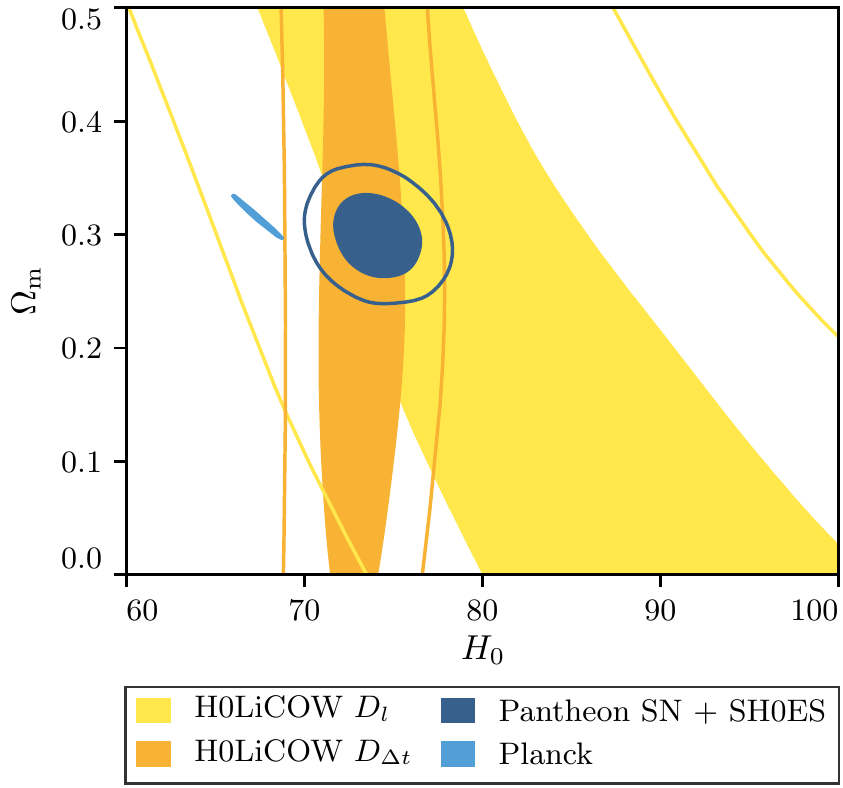}
\caption{ \label{Fig:SLSNLCDMComparison}
The joint marginalized posterior distribution of the two flat $\Lambda$CDM parameters, $\Omega_m$ and $H_0$, as obtained from the calibrated Pantheon SN sample, H0LiCOW  time delays, and {\it Planck} CMB observations.
The filled contour corresponds to the 68\% C.L. region while the lines  shows the 95\% C.L. region.
}
\end{figure} 
As a further check we fit to the  measurements (both individuallhy and jointly) the flat $\Lambda$CDM model finding good agreement with the results in~\cite{Wong:2019kwg}. 
Since we do not have access to the joint time delay and lens distance posterior we cannot combine the two measurements.
We therefore show in Fig.~\ref{Fig:SLSNLCDMComparison} the joint posterior on $\Omega_m$ and $H_0$ for the time delay and lens distance measurements.
The time delay result measures $H_0=73.1\pm 1.8$, in good agreement with~\cite{Wong:2019kwg}.
As a reference we show the results from the fully calibrated Pantheon sample, highlighting their agreement with H0LiCOW within  the $\Lambda$CDM model, and the results from {\it Planck} CMB measurements.

\section{ Gaussian process regression on the Supernova data}\label{App:SN_GP}

In this appendix we  describe the details of the SN Gaussian process (GP) regression~\cite{Rasmussen:2005:GPM:1162254} that allows for the computation of the predicted time delay distances.

The goal of our GP construction is to build, starting from the full SN catalog, a continuous version of the SN sample that is in quantitative agreement with the binned data that the Pantheon collaboration provides.

The first step in the construction of the GP is to define the time scale that is used.
The binned SN data points are not equally spaced in redshift
~\cite{Scolnic:2017caz,Brout:2018jch}. 
In order to ensure that the GP recovers the relative weights of the binned data points as solely determined by the data covariance, and that it does not introduce artefacts that depend on the relative distance between the points, we find a unique mapping between a uniform redshift spacing and the redshift values of the binned Pantheon data. 
This effectively defines a new time coordinate in which the binned SN sample is equispaced and this coordinate is then used to define the GP.

In order to reliably compare the estimated time delay distances, we need to make sure that the GP has enough freedom to describe all variations in the measured distance modulus from the Pantheon data.

The fluctuations in the SN binned data points are small compared to overall variations of the distance-redshift relation. To help the GP in recovering them we subtract from the data the best fit $\Lambda$CDM distance modulus that is added back to the GP predictions. 
Notice that this is just a convenient choice to work with data points that are scattered around zero.

The degree of flexibility of the Gaussian process depends on the choice of the kernel function describing relative correlation between different points with different time coordinates. 
We consider three different kernels to interpolate the full SN catalog:
the exponential squared kernel; 
the Matern kernels of order $\nu = 9/2$;
the rational quadratic kernel.
The first two kernels have two free parameters describing the strength and the characteristic time scale of the correlation.
The last kernel, in addition to these two parameters, has a parameter, the decay exponent, that describes the relative weighting of the large and small scale variations.

To all these kernels we add a constant offset that we need to accurately reproduce the correlation between the SN measurements induced by the shared amplitude calibration, as in Eq.~\eqref{Eq:SNCalibratedCovariance}.

The kernel parameters are usually selected by maximizing the marginal likelihood (i.e. the evidence) of the observed data predicted using the GP.
This method is, however, not guaranteed to converge to a unique working solution.
We find that, in our case, this does not provide a good solution for the GP.

To select the best kernel parameters we then minimize the
Kullback-Leibler (KL) divergence~\cite{kullback1951information} between the GP prediction and the binned Pantheon data.

The KL divergence represents the information difference in going from the GP to the full binned data and, in this case, serves as a measure of how much the two distributions differ.
The KL divergence, $D(P_{\rm d}||P_{\rm gp})$, for the two Gaussian distributions, $P_{\rm gp}$ and $P_{\rm d}$, is given by:
\begin{align} \label{Eq:KLOptimization}
D(P_{\rm d}||P_{\rm gp}) \equiv \frac{1}{2 \ln 2} \Bigg[ & (y_d - y_{gp})^T \Sigma_d^{-1} (y_d - y_{gp}) \nonumber \\
    & -\ln \frac{{\rm det} \  \Sigma_{gp} }{{\rm det} \  \Sigma_d} - d + {\rm tr}(\Sigma_{gp} \Sigma_d^{-1}) \Bigg] \,, \nonumber \\
\end{align}
where $y_d$ and $y_{gp}$ are the distance modulus residuals from the binned SN data and predictions from the GP at a particular set of kernel parameters, $\Sigma_d$ is the covariance of the binned data, $\Sigma_{gp}$ is the Gaussian process predicted covariance conditioned on the full Pantheon data and $d$ is the number of the binned data points.

We find that this optimization problem, for all the kernels that we consider, has always a unique and well defined solution.

This also allows us to compare the performances of different kernels in matching the binned SN data.
We find that using the rational quadratic kernel leads to the lowest KL divergence value, by a marginal amount, and hence we use this as our fiducial choice of kernel to produce the main results of this paper. 

\begin{figure}[htbp]
\centering
\includegraphics[width=\columnwidth]{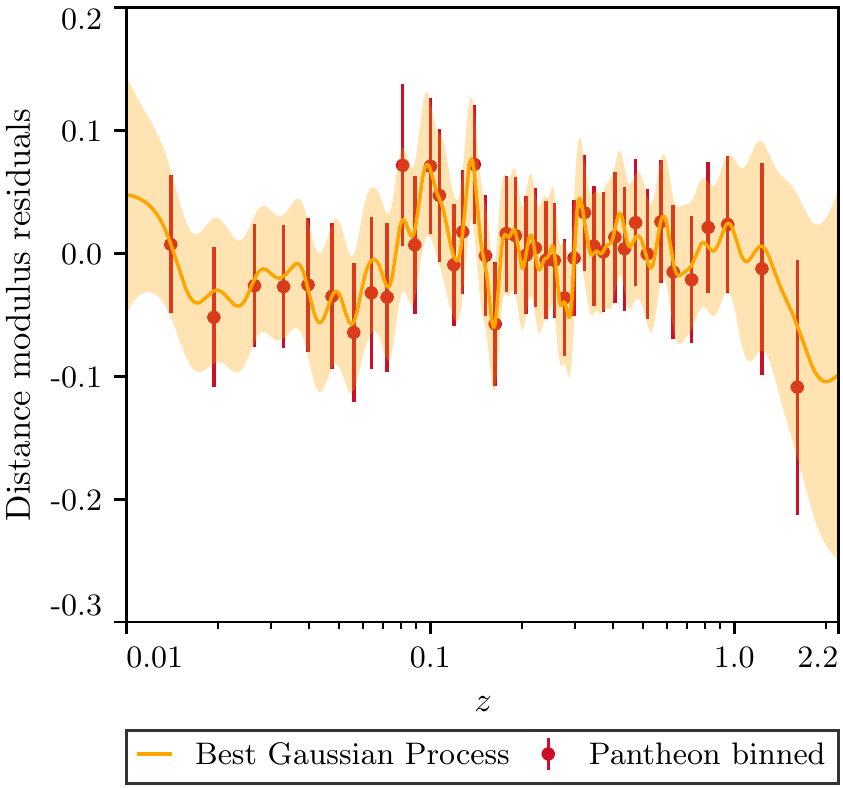}
\caption{ \label{Fig:AppBinnedGPComparison}
Comparison of the binned Pantheon data and the predictions from the GP on the full SN sample, with the best available kernel and kernel parameters. 
The continuous line represents the predicted GP mean while the color band represents the GP predicted uncertainty. 
In this figure the best fit $\Lambda$CDM model has been subtracted to highlight the features in the binned SN sample.
}
\end{figure} 
In Fig.~\ref{Fig:AppBinnedGPComparison} we compare the results obtained from the GP regression of the full SN sample that we discussed, for the best matching kernel, to the binned Pantheon measurements.
As we can see this results in excellent agreement between the two on the reconstructed distance modulus, and its variance, that preserves all the features of the binned data points.
We highlight that our goal is to preserve all features in the distance measurements that are present in the binned sample rather than smoothing it further.

\begin{figure}[htbp]
\centering
\includegraphics[width=\columnwidth]{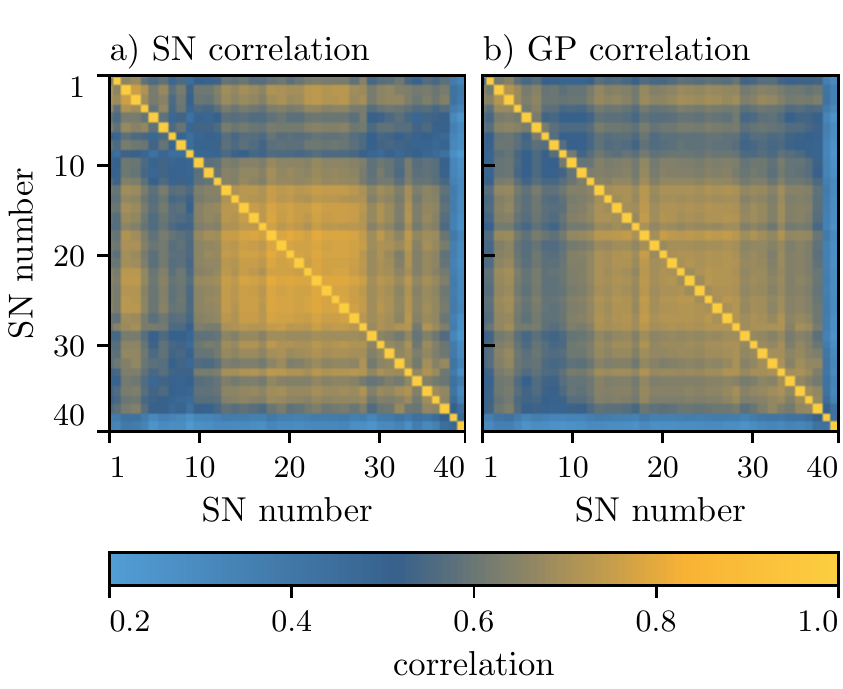}
\caption{\label{Fig:AppBinnedGPCorrelationComparison}
The correlation matrix of the binned calibrated Pantheon distance modulus compared to the correlation matrix predicted by the GP regression of the full SN sample.
}
\end{figure} 
We also test that the Gaussian process is recovering the correlation between different data points as in the binned SN sample. These correlations are induced by the common shared calibration of the SN and joint correction of systematic effects, as discussed in~\cite{MRinprep}.
In Fig.~\ref{Fig:AppBinnedGPCorrelationComparison} we compare the binned Pantheon correlation matrix with the GP predicted correlation. As we can see these correlations are not negligible, as discussed in~\cite{MRinprep}, and are faithfully reproduced by the GP regression of the full SN sample.

\begin{figure}[htbp]
\centering
\includegraphics[width=\columnwidth]{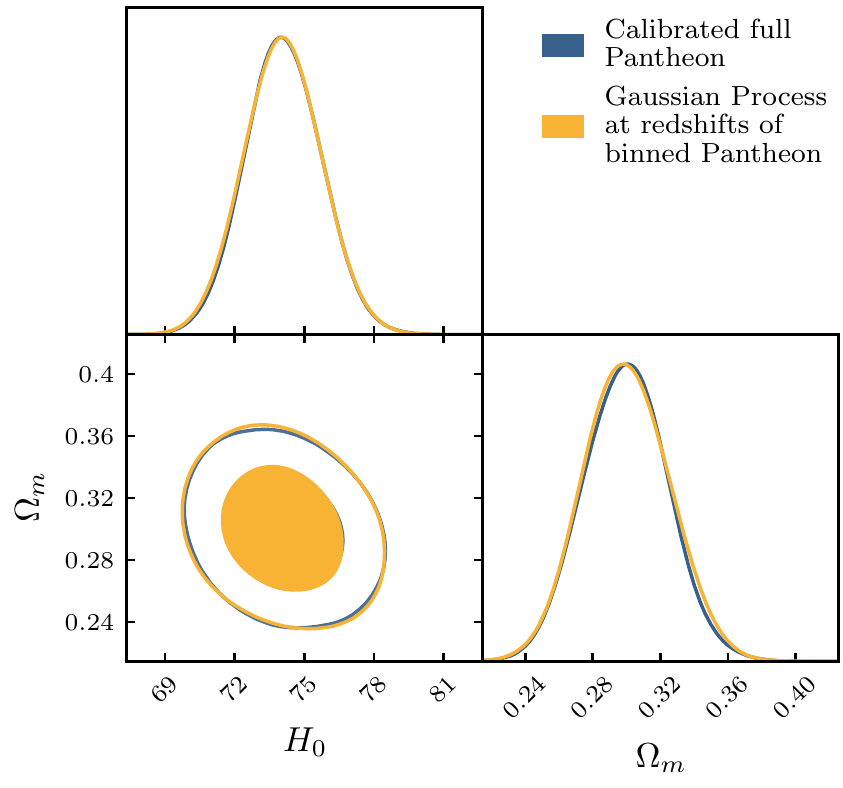}
\caption{ \label{Fig:AppGPLCDMParamPosterior}
The comparison of the posterior of the $\Lambda$CDM parameters as obtained by fitting the full Pantheon data set and by fitting the GP prediction evaluated at the redshifts of the binned SN sample.
The filled contour corresponds to the 68\% C.L. region while the continuous contour shows the 95\% C.L. region.
This ensures that the GP is a faithful representation of the SN data and is not introducing any subtle changes in the predictions across redshift.
}
\end{figure} 
To further test the fidelity of the GP regression we want to ensure that the kernel smoothing is not introducing subtle inconsistencies in either the mean and the covariance of the predicted measurements.
These include over smoothing the data that would result in a wrongly overestimated data constraining power.
To do so we take the GP predicted mean and covariance at the redshifts of the binned Pantheon data set and we fit them with the $\Lambda$CDM model as if they were the true data points.
In Fig.~\ref{Fig:AppGPLCDMParamPosterior} we show the excellent agreement between the full Pantheon posterior and the posterior obtained fitting the GP predictions.
Furthermore we have checked that, if we were to take more data points, the results will still perfectly agree with the full data set, meaning that the GP is not introducing spurious data constraining power.

\begin{figure*}[tbp]
\centering
\includegraphics[width=\textwidth]{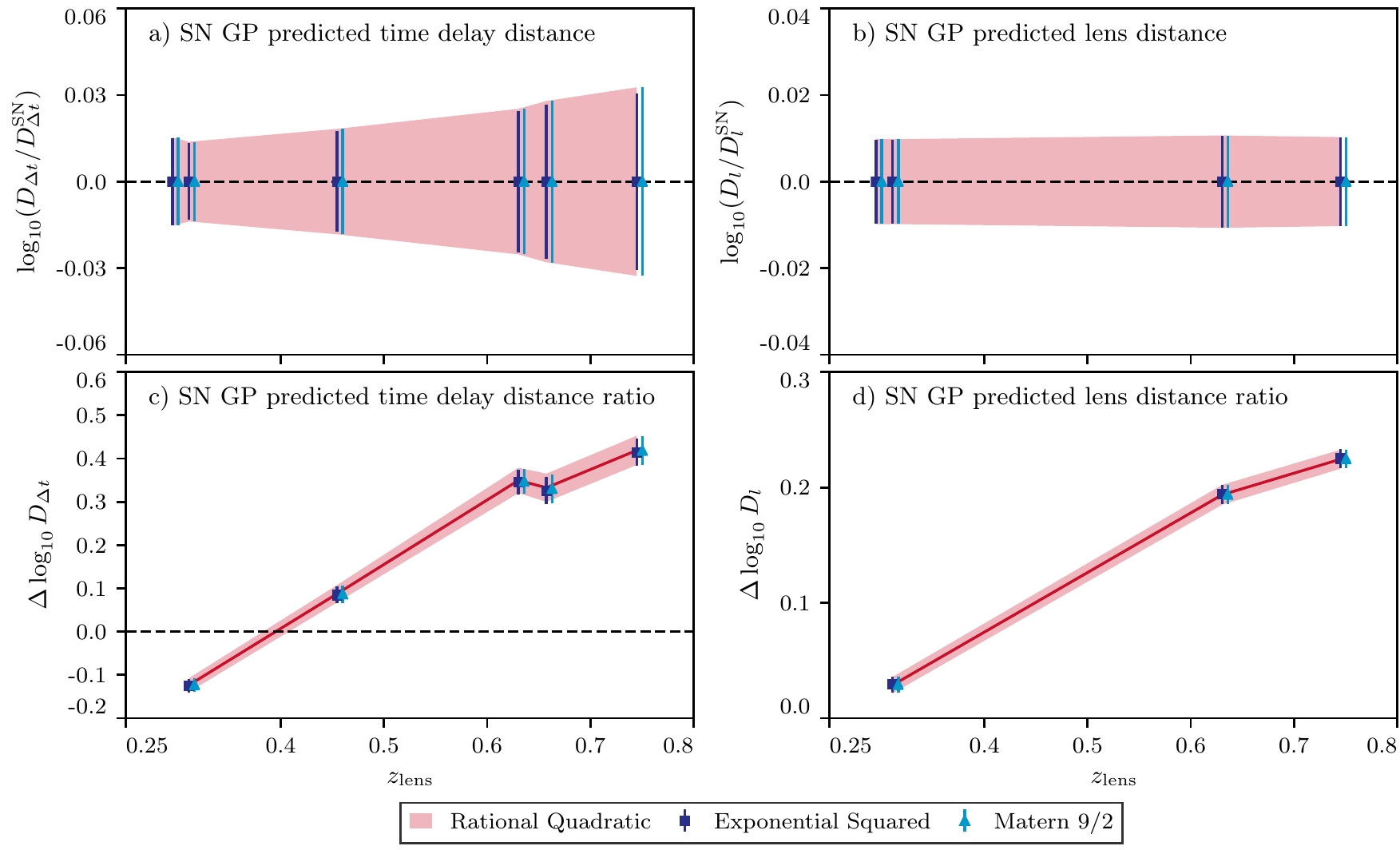}
\caption{ \label{Fig:AppDifferentKernelTest}
Comparison of the end to end results, as in Fig. \ref{Fig:DistanceComparison} and Fig. \ref{Fig:DistanceRatioComparison}, 
for the different GP kernels that we consider. The red solid line and red band shows the results for our fiducial rational quadratic kernel. We shift the data points of other two kernels in redshift by a small amount for visual clarity. 
}
\end{figure*}

To further ensure the stability of our results to the specific choice of the GP kernel we run the end to end analysis for all the three kernels that we consider.
In Fig.~\ref{Fig:AppDifferentKernelTest} we report the results, as in Fig.~\ref{Fig:DistanceComparison} and Fig.~\ref{Fig:DistanceRatioComparison}, 
clearly showing quantitative agreement between different kernels.

\begin{figure}[tbp]
\centering
\includegraphics[width=\columnwidth]{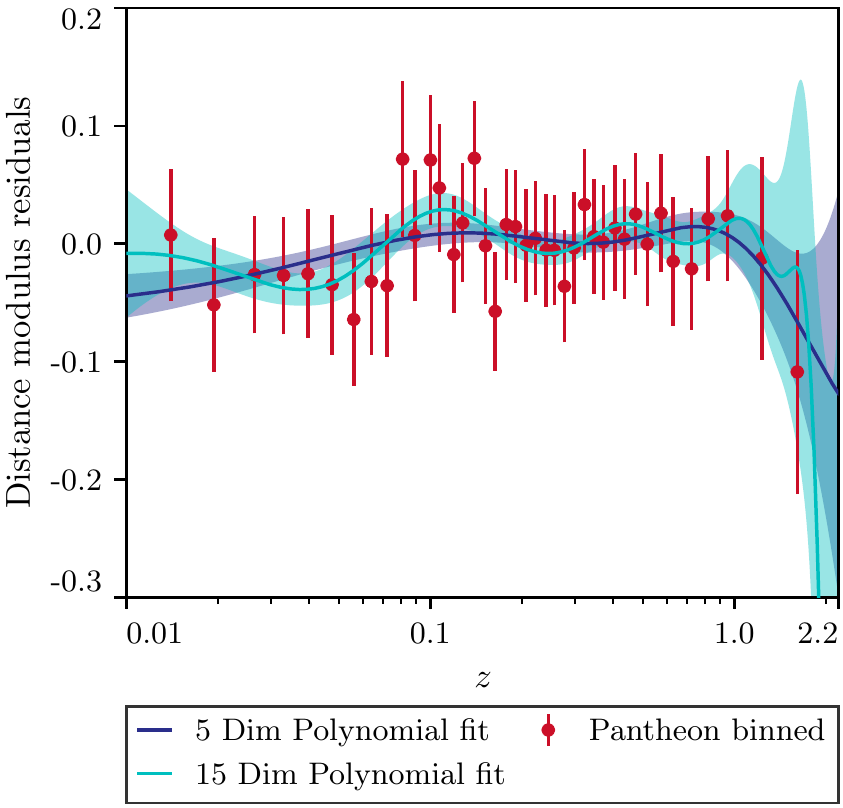}
\caption{ \label{Fig:AppPolynomianTest}
Comparison of the binned Pantheon data and the polynomial fit of degree 5 and 15.
The continuous line represents the mean of the polynomial fit while the color band represents the predicted uncertainty. 
Comparing this to Fig.~\ref{Fig:AppBinnedGPComparison} we see that using polynomial fit results in wrong error predictions across redshift. 
This highlights the need for a more sophisticated GP regression technique to recover unbiased estimate of covariance between the points.
In this figure the best fit $\Lambda$CDM model has been subtracted to highlight the features in the binned SN sample.
}
\end{figure} 
Finally we can compare our predictions to the case of fitting the binned distance modulus with a polynomial in log-redshift.
We consider a 5-degree and 15-degree polynomial and show the results of the fit in Fig.~\ref{Fig:AppPolynomianTest}.
Qualitatively the mean behavior reproduces the results of the GP for the two different degrees, with the higher order polynomial preserving more details of the data.
The covariance of the polynomial fit is, however, largely misestimating the covariance of the true data points.
In particular the low degree polynomial case is underestimating the covariance so that using this technique would clearly give overly constrained results.
On the other hand the high order polynomial is overestimating the covariance at the end points while underestimating it toward the center of the redshift distribution.
This is because, outside of the radius of convergence of the polynomial expansion, polynomials grow unbounded. This is then responsible for over constrained results inside the convergence radius and under constrained results outside of it. 
We have also compared our results with some of the polynomial models considered in~\cite{Arendse:2019hev}, by changing the interpolation time scales and we reach similar conclusions.

All these issues are not present in our GP construction that overall gives a more faithful representation of the data.

\section{Tests with biased fake H0LiCOW Data}\label{App:SL_Bias}

\begin{figure*}[tbp]
\centering
\includegraphics[width=\textwidth]{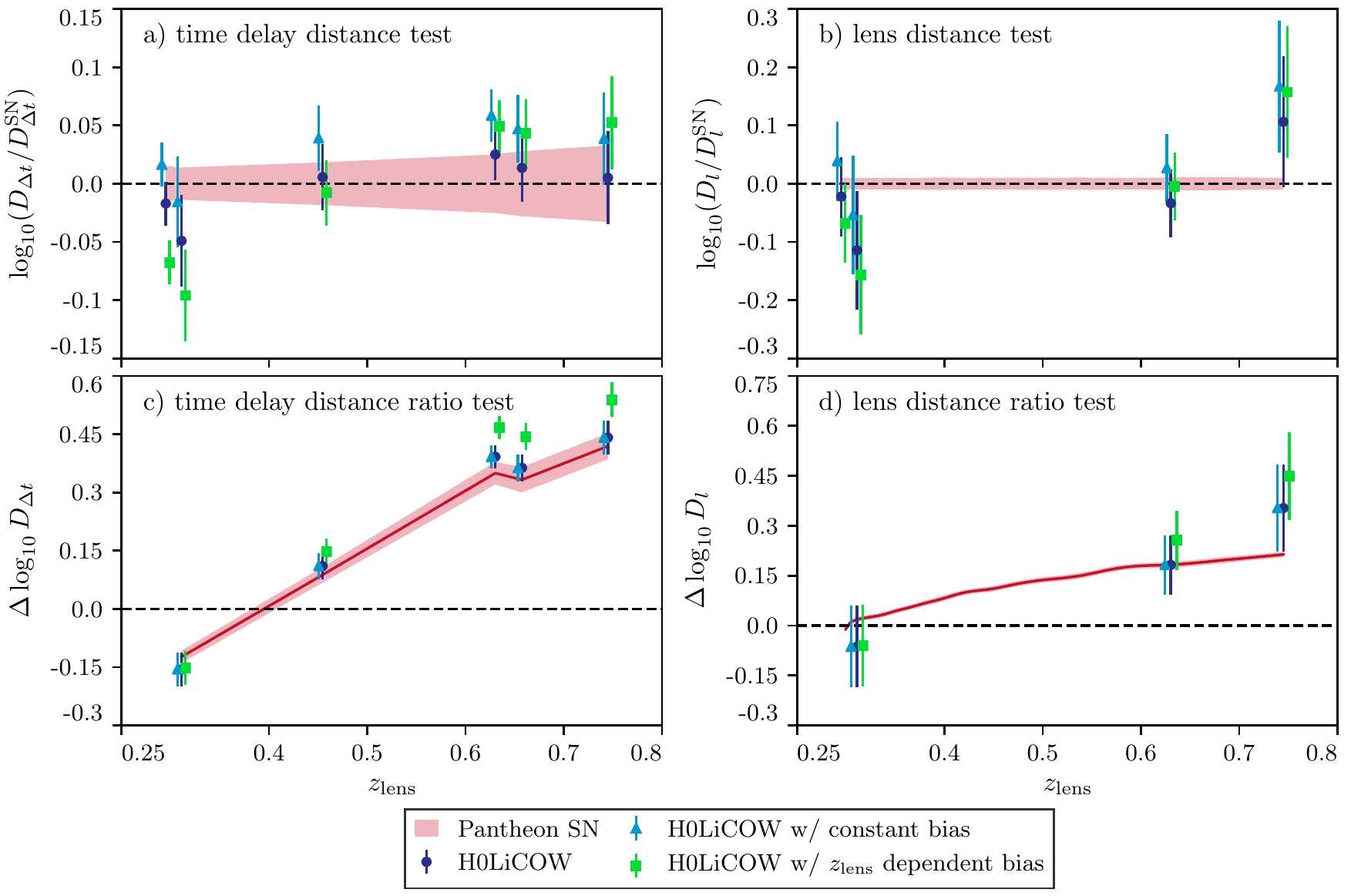}
\caption{ \label{Fig:BaisedDistanceComparison}
Comparison of the measured  H0LiCOW measured time delay distances $D_{\Delta t}$ (a) and lens distances $D_\lens$ (b) with the Pantheon SN predicted values. These are similar to Fig. \ref{Fig:DistanceComparison} but with two sets of additional points, corresponding to  biases in the H0LiCOW distances. The constant 8\% bias in $D_{\Delta t}$ and 15\% bias in $D_{\lens}$ shifts the inferred $H_0$ from H0LiCOW to match {\it Planck} (see Fig.~\ref{Fig:SLSNLCDMComparison}). This bias results in increased tension with respect to SN predictions as detailed in Table~\ref{Tab:SLBiasComp}. Since this is a constant shift, it does not affect the distance ratio tests shown in the lower panels. The green points with square markers show fake measurements with redshift dependent bias parameterized by $0.5(z_{\rm lens}-z_{\rm mean})$, where $z_{\rm mean}$ is the mean redshift of the lenses. This results in a similar mean bias as in the upper panels. It is evident that all the tests shown here result in increased tension between the two datasets. 
}
\end{figure*}

In this section we detail the tests on fake H0LiCOW data created by biasing  the measured $D_{\Delta t}$ and $D_{\lens}$ by a specified percentage.
We change both the mean and covariance of the measurements to keep the signal to noise ratio of the measurements constant.
We then analyze the consistency of these fake data sets with SN and {\it Planck} data in both a model independent way and within the $\Lambda $CDM model. 
We show the comparison of these shifted data with the original data and SN predictions in Fig.~\ref{Fig:BaisedDistanceComparison}. 
We find that a +8\% shift in $D_{\Delta t}$ and +15\% shift in $D_{\lens}$ predictions would resolve the $H_0$ tension between {\it Planck} and H0LiCOW, but would result in a 2.85$\sigma$ tension between SN and H0LiCOW on the determination of $H_0$ (the tension increases to over 3$\sigma$ for a 10\% shift in $D_{\Delta t}$. )
We can also do this comparison in a model independent way by comparing the amplitude of the shifted H0LiCOW data points with the SN predictions. This comparison results in a tension of 1.45$\sigma$ over the amplitude of $\log_{10} D_{\Delta t}$ and 0.6$\sigma$ in $\log_{10} D_{\lens}$ with respect to SN measurements.

A negative bias of -8\% for time delay distances and -15\% for lens distances would decrease the agreement of H0LiCOW with both {\it Planck} and SN. 
In particular, considering time delay distances, the $H_0$ determination would be in 5.8$\sigma$ tension with {\it Planck} and 2.2$\sigma$ tension with SN.
The model independent amplitude test would report a 1.8$\sigma$ tension between SN and H0LiCOW on both time delay and lens distance measurements.

Notice that  these tests would leave the model independent shape test invariant since we are considering a multiplicative bias that is the same for all H0LiCOW measurements.

We also test a redshift dependent bias parameterized by $0.5(z_{\rm lens}-z_{\rm mean})$  where $z_{\rm mean}$ is the mean redshift of the lenses. This results in a mean bias amplitude of 8\% in $D_{\Delta t}$ and 15\% bias in $D_{\lens}$, consistent with the constant bias case. Unknown measurement systematics or  incorrect lens modelling are likely to affect different systems differently, so they  could show up as an unexpected redshift dependence. 
These problems are not guaranteed to show up as a discrepancy in the determination of cosmological parameters, so our tests provide an independent check.  We get a 2.7$\sigma$ tension between these fake measurements and SN predictions for the time delay distance test and a 2.5$\sigma$ tension for time delay distance ratio test. The values for the lens distance tests are given in Table~\ref{Tab:SLBiasComp}. 

Finally, we test the possibility that the random error is underestimated by inflating the covariance of the measurements  to give a 10\% uncertainty on the resulting $H_0$. This could represent a scenario where lens mass modeling adds to the scatter.  
We  inflate the error bar on $D_{\Delta t}$ measurements by a factor four (we do not consider $D_\lens$  since these already result in a $10\%$ determination of $H_0$). 
In this case, comparing SN predictions to the fake data gives a probability of disagreement of $0.05\%$ for the full test and $0.2\%$ for the shape test.
Both these results  therefore signal a statistically significant excess confirmation.

We have also checked the alternative value of $H_0$ obtained using the TRGB~\cite{Freedman:2019jwv} calibration of local SN. Since their value of $H_0$ is lower, we get slightly higher tension with strong lensing though not at high significance given the larger error bars in the determination of $H_0$.
The global test of time delay and lens distance is in agreement at the $58.8\%$ and $45.7\%$ confidence level respectively.
The amplitude test gives slightly higher tension with a probability of agreement of $25.1\%$ and $37.7\%$ respectively for time delay and lens distances.
All of these results are summarized in the Table~\ref{Tab:SLBiasComp}.
\begin{table*}[ht!]
\begin{ruledtabular}
\begin{tabular}{ l c c c c c c c }
\textrm{Data Vector} & 
\begin{tabular}{@{}c@{}}$H_0$ \\ test ($\sigma$) \end{tabular} &
\begin{tabular}{@{}c@{}}full $D_{\Delta t}$\\ test ($\sigma$) \end{tabular} &
\begin{tabular}{@{}c@{}}$D_{\Delta t}$ Amplitude\\ test ($\sigma$) \end{tabular} &
\begin{tabular}{@{}c@{}}$D_{\Delta t}$ Shape\\ test ($\sigma$) \end{tabular} &
\begin{tabular}{@{}c@{}}full $D_{\lens}$\\ test ($\sigma$) \end{tabular} &
\begin{tabular}{@{}c@{}}$D_{\lens}$ Amplitude\\ test ($\sigma$) \end{tabular} &
\begin{tabular}{@{}c@{}}$D_{\lens}$ Shape\\ test ($\sigma$) \end{tabular}  \\
\hline \hline
Original  & 0.4     & 0.2 & 0.1 & 0.3 & 0.5 & 0.3 & 0.6 \\
TRGB Calibration & 1.0 & 0.5 & 1.1 & 0.3 & 0.7 & 0.8 & 0.6 \\
Constant Bias & 2.9   & 0.6   & 1.5 & 0.3 & 0.5 & 1.0 & 0.6  \\
Redshift-dependent Bias & $0.3$   & $2.6$ & $0.2$ & $2.5$ & $1.1$ & $0.4$ & $1.3$ \\
Increased Uncertainty & 0.1 & 0.0006 & 0.2 & 0.002 & $-$ & $-$ & $-$ \\
\end{tabular}
\end{ruledtabular}
\caption{\label{Tab:SLBiasComp}
Results of our tests on the variations of the original comparison of SN and H0LiCOW data, as described in the first column. We quote the recovered tension (in terms of effective numbers of $\sigma$, given by $n_\sigma^{\rm eff} \equiv \sqrt{2}{\rm Erf}^{-1}(P)$, where $P$ is the probability of disagreement and ${\rm Erf}$ is the error function) in various tests with respect to the inferences from GP regression on SN data.  
Note that the small values in the last row, which corresponds to increasing the $H_0$ uncertainty to 10\%, indicate a statistically significant excess confirmation.
The datavectors corresponding to the  TRGB Calibration, constant bias and redshift-dependent bias result in a mean bias of 8\% in $D_{\Delta t}$ and 15\% in $D_{\lens}$ compared to the original measurements (chosen to resolve the tension with {\it Planck}). The data vectors corresponding to the original, constant bias and redshift dependent bias cases are also shown in Figure~\ref{Fig:BaisedDistanceComparison}.
}
\end{table*}

\bibliographystyle{apsrev4-1}
\bibliography{biblio}

\end{document}